\documentstyle[12pt,psfig]{article}
\begin{document}

\title{
Hypernucleus formation and strangeness production in proton-nucleus reactions
\footnote{Supported by  Forschungszentrum J\"ulich and the
Polish Committee for Scientific Research }}
\author{ W. Cassing$^2$, Z. Rudy$^1$,
         L. Jarczyk$^1$, B. Kamys$^1$, \\
       P. Kulessa$^1$, O. W. B. Schult$^3$, A. Sibirtsev$^2$,
       A. Strza\l kowski$^1$  \\  \\
        $^1$ Institute of Physics, Jagellonian University, \\
        Reymonta 4, PL-30059 Cracow, Poland \\
     $^2$ Institut f\"ur Theoretische Physik, \\
        Heinrich-Buff-Ring 16, D-35392 Giessen, Germany \\
        $^3$ Experimentelle Kernphysik II, Forschungszentrum J\"ulich, \\
        D-52425 J\"ulich, Germany }
\date{}
\maketitle

\begin{abstract}
We study the production of $\Lambda$ hyperons in $p + A$ reactions on
the basis of a BUU transport approach from 1.1 to 1.9 GeV and evaluate
the properties of the hypernuclei produced in particular with respect
to their momentum distribution in the laboratory frame. Due to elastic
$\Lambda N$ scattering large cross sections for the production of heavy
hypernuclei in the order of 100 - 400 $\mu b$ are predicted for p + U
at 1.5 - 1.9 GeV laboratory energy. Whereas the $K^+ Y$ production
channels are expected to be only slightly modified in the nuclear
medium, the antikaon production should be enhanced substantially due to
large attractive $K^-$ selfenergies in dense matter. We predict an
enhancement of the inclusive $K^-$ yield in p + $^{208}$Pb collisions
of a factor of $\approx$ 10 at 2 GeV   laboratory energy.
\end{abstract}


\section{Introduction}

The investigation of strangeness in hadrons and nuclei has become an
exciting and challenging field of research in the last years \cite{9}.
Here, the open questions reach from the  spectroscopy of hypernuclei to
the formation of strange hadronic matter in neutron stars or
ultrarelativistic nucleus-nucleus collisions. Whereas earlier studies
were devoted to the production aspects of hyperons and strange mesons,
nowadays, the interest is moving towards the properties of strange
particles in a  nuclear environment.

$\Lambda$-hypernuclei are especially well suited for in-medium
strangeness investigations since the $\Lambda$-hyperon has a long
lifetime as compared to the nuclear scale.  The 'free'
 $\Lambda$-hyperon decay is purely mesonic ($\Lambda \rightarrow N
\pi$) whereas in the nuclear medium also nonmesonic decay channels
($\Lambda N \rightarrow N N$) are possible.  The competition between
these decay modes is expected to provide information about the
hypernuclear structure (e.g. the $\Lambda$ and pion selfenergies in the
nuclear medium) \cite{19}.  Since the lifetime of the hypernucleus
depends on the corresponding widths of these decays , i.e.  $\tau$ =
 $\hbar$ / [ $\Gamma_{mes}$ + $\Gamma_{nonmes}$], it is of considerable
interest to measure the lifetime of hypernuclei as a function of their
mass, particularly for heavy systems, where $\Gamma_{mes}$ is strongly
 suppressed due to Pauli blocking \cite{Rudy} such that
$\Gamma_{nonmes}$ is directly related to $\tau^{-1}$.

This particular question has been investigated with antiproton beams on
 $^{209}$Bi and $^{238}$U targets in \cite{3}. However, the lifetimes
of $^{209}_{\Lambda}$Bi and $^{238}_{\Lambda}$U from this experiment
are quite different contrary to simple phase-space expectations.
Alternatively, one might investigate the production of heavy
hypernuclei by means of proton + nucleus reactions.  In fact, the
recent studies on the (p , K$^{+}$ ) reaction confirm a quite
substantial production of associated $\Lambda$-hyperons \cite{13,22}
leading to production cross sections for $\Lambda$-hypernuclei in the
order of a few 100 $\mu b$ for p + Pb at 1.5 - 1.9 GeV
\cite{Rudy,Rudy2}.

Apart from the lifetime of the $\Lambda$ hyperon in the medium also the
kaon and antikaon properties should change due to interactions with the
nuclear environment \cite{GB1,Kaplan,waas}. Whereas the $K^+$ potential
is expected to be slightly repulsive in the medium, the $K^-$ meson
should see a sizeable attractive potential in dense matter such that
their production should be enhanced at finite baryon density. In case
of heavy-ion collisions a first exploratory study has been performed by
Li, Ko and Fang \cite{Fang} with the result that large attractive $K^-$
potentials are needed to explain the experimental spectra from
\cite{Schro} for Ni + Ni at 1.85 AGeV.  A more systematic transport
analysis - including all possible production channels - on this
question has been performed in Refs. \cite{Ca97,Brat97} in comparison
to the recent data from the KaoS Collaboration \cite{kaos}. In fact,
the experimental data indicate an attractive $K^-$ potential of about
-100 MeV at normal nuclear matter density ($\rho_0 \approx$ 0.16
fm$^{-3}$) whereas the $K^+$ potential at density $\rho_0$ is in the
range between 0 and +30 MeV \cite{Ca97,Brat97}. Such effects should
also be seen and probed experimentally in p + A reactions at
subthreshold energies.

In continuation of our earlier work \cite{Rudy,Rudy2} we here present a
transport (BUU) analysis of proton + nucleus collision events,
including the production channels $pN \rightarrow N Y K^+$ and $\pi N
\rightarrow Y K^+$ (Y = $\Lambda, \Sigma$), while nonperturbatively
taking into account the rescattering of the hyperons with nucleons and
the hyperon propagation in the nuclear mean field.

\section{Ingredients of the transport approach}

In this contribution we perform our analysis along the line of the
HSD\footnote{Hadron String Dynamics} approach~\cite{Ehehalt} which is
based on a coupled set of covariant transport equations for the
phase-space distributions $f_{h} (x,p)$ of hadron
$h$, i.e.
\begin{eqnarray}  \label{g24}
\lefteqn{\left\{ \left( \Pi_{\mu}-\Pi_{\nu}\partial_{\mu}^p U_{h}^{\nu}
-M_{h}^*\partial^p_{\mu} U_{h}^{S} \right)\partial_x^{\mu}
+ \left( \Pi_{\nu} \partial^x_{\mu} U^{\nu}_{h}+
M^*_{h} \partial^x_{\mu}U^{S}_{h}\right) \partial^{\mu}_p
\right\} f_{h}(x,p) } \nonumber \\
&& = \sum_{h_2 h_3 h_4\ldots} \int d2 d3 d4 \ldots
 [G^{\dagger}G]_{12\to 34\ldots}
\delta^4(\Pi +\Pi_2-\Pi_3-\Pi_4 \ldots )  \nonumber\\
&& \times \left\{ f_{h_3}(x,p_3)f_{h_4}(x,p_4)\bar{f}_{h}(x,p)
\bar{f}_{h_2}(x,p_2)\right.  \nonumber\\
&& -\left. f_{h}(x,p)f_{h_2}(x,p_2)\bar{f}_{h_3}(x,p_3)
\bar{f}_{h_4}(x,p_4) \right\} \ldots\ \ .
\end{eqnarray}
In Eq.~(\ref{g24}) $U_{h}^{S}(x,p)$ and $U_{h}^{\mu}(x,p)$ denote the
real part of the scalar and vector hadron selfenergies, respectively,
while $[G^+G]_{12\to 34\ldots} \delta^4 (\Pi
+\Pi_2-\Pi_3-\Pi_4\ldots )$ is the 'transition rate' for the process
$1+2\to 3+4+\ldots$ which is taken to be on-shell in the
semiclassical limit adopted. The hadron quasi-particle properties in
(\ref{g24}) are defined via the mass-shell constraint
\begin{equation}   \label{g25}
\delta (\Pi_{\mu}\Pi^{\mu}-M_{h}^{*2} ) \ \ ,
\end{equation}
with effective masses and momenta (in local Thomas-Fermi approximation)
given by
\begin{eqnarray}\label{g26}
M_{h}^* (x,p)&=&M_h + U_h^{{S}}(x,p) \nonumber \\
\Pi^{\mu} (x,p)&=&p^{\mu}-U^{\mu}_h (x,p)\ \ ,
\end{eqnarray}
while the phase-space factors
\begin{equation}
\bar{f}_{h} (x,p)=1 \pm f_{{h}} (x,p)
\end{equation}
are responsible for fermion Pauli-blocking or Bose enhancement,
respectively, depending on the type of hadron in the final/initial
channel. The dots in Eq.~(\ref{g24}) stand for further contributions to
the collision term with more than two hadrons in the final/initial
channels. The transport approach (\ref{g24}) is fully specified by
$U_{h}^{S}(x,p)$ and $U_{h}^{\mu}(x,p)$ $(\mu =0,1,2,3)$, which
determine the mean-field propagation of the hadrons, and by the
transition rates $G^\dagger G\,\delta^4 (\ldots )$ in the collision
term, that describes the scattering and hadron production/absorption
rates.

The scalar and vector mean fields $U_{h}^{S}$ and $U^\mu_{h}$ for
nucleons are taken from Ref.~\cite{Ehehalt}; the hyperon mean fields
are assumed to 2/3 of the nucleon potentials. In the present approach,
apart from nucleons, $\Delta, N(1440)$, $N(1535)$, $\Lambda, \Sigma$
with their isospin degrees of freedom, we propagate explicitly pions,
$K^+, K^-$, and $\eta$'s and assume that the pions as Goldstone bosons
do not change their properties in the medium; we also discard
selfenergies for the $\eta$-mesons which have a minor effect on the
$K^+, K^-$ dynamics.  The kaon and antikaon potentials, however, have
to be specified more explicitly.

\subsection{$K^+, K^-$ selfenergies}

There are a couple of models for the kaon and
antikaon selfenergies~\cite{GB1,Kaplan,waas},
which differ in the actual magnitude of the selfenergies, however,
agree on the relative signs for kaons and antikaons. Thus in line with
the kaon-nucleon scattering amplitude the $K^+$ potential should be
slightly repulsive at finite baryon density whereas the antikaon should
see an attractive potential in the nuclear medium. Without going into a
detailed discussion of the various models we adopt the more practical
point of view, that the actual $K^+$ and $K^-$ selfenergies are unknown
and as a guide for our analysis use a linear extrapolation of the form,
\begin{equation}
\label{kmass}
m^*_K(\rho_B) = m_K^0 \left(1 + \alpha \frac{\rho_B}{\rho_0}\right),
\end{equation}
with $\alpha \approx $ -0.2 for antikaons and $\alpha \approx 0.06$ for
kaons (or $\alpha$ = 0 for the bare kaon).  Our choice ($\alpha \approx
$ -0.2) leads to a fairly reasonable reproduction of the antikaon mass from
Ref. \cite{Kaplan} (thin solid line in Fig. 1) and the recent
results from Waas, Kaiser and Weise \cite{waas} (thick solid line in
Fig. 1).  We note that the dropping of the antikaon mass is associated
with a corresponding scalar energy density in the baryon/meson
Lagrangian, such that the total energy-momentum is conserved during the
heavy-ion collision (cf.~\cite{Ehehalt}).
\vbox to 7.5cm {
\psfig{figure=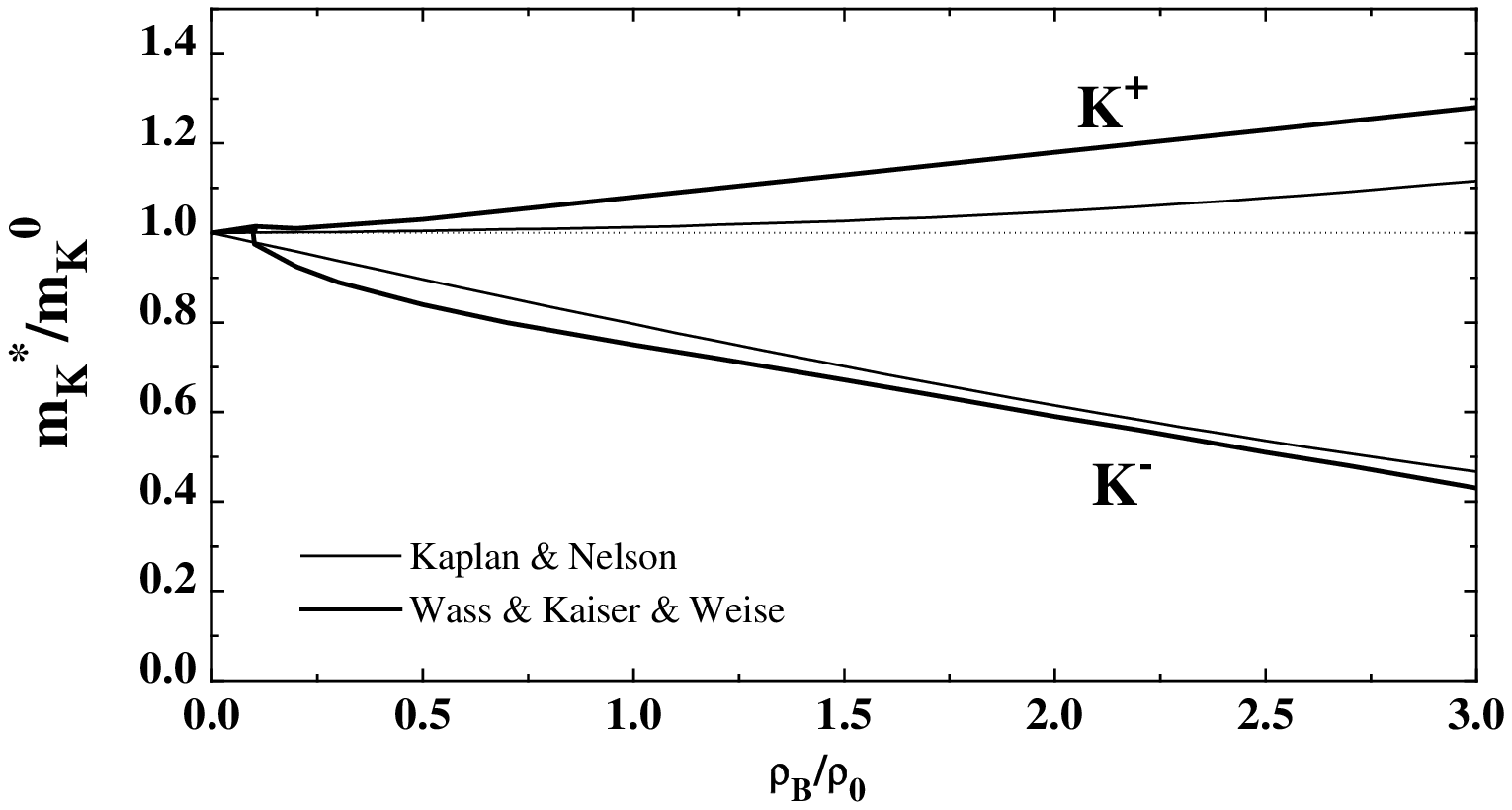,width=15cm,height=22cm}}
\\
\noindent
{\bf Fig. 1:}
The  $K^+, K^-$ mass as a function of the baryon density in units
of $\rho_0 \approx 0.16 fm^{-3}$ according to Kaplan and Nelson
\cite{Kaplan} (thin solid lines) and Waas, Kaiser and Weise
\cite{waas} (thick solid lines).

\subsection{Perturbative treatment of strangeness production}
The calculation of 'subthreshold' particle production is described in
detail in Ref.~\cite{24} and has to be treated perturbatively in the
energy regime of interest here due to the small cross sections
involved. Since we work within the parallel ensemble algorithm, each
parallel run of the transport calculation can be considered
approximately as an individual reaction event, where binary reactions
in the entrance channel at given invariant energy $\sqrt{s}$ lead to
final states with 2 (e.g. $K^+ Y$ in $\pi B$ channels), 3 (e.g. for
$K^+ YN$ channels in $BB$ collisions) or 4 particles (e.g. $K\bar{K}NN$
in $BB$ collisions) with a relative weight $W_i$ for each event $i$
which is defined by the ratio of the production cross section to the
total hadron-hadron cross section\footnote{The actual final states are
chosen by Monte Carlo according to the 2, 3, or 4-body phase space.}.
We thus dynamically gate on all events where a $K^+ Y$ or $K^+ K^-$
pair is produced initially.  Each strange hadron then is represented by
a testparticle with weight $W_i$ and propagated according to the
Hamilton equations of motion. Elastic and inelastic reactions with
pions, $\eta$'s or nonstrange baryons are computed in the standard
way~\cite{24} and the final cross section is obtained by multiplying
each testparticle with its weight $W_i$.  In this way one achieves a
realistic simulation of the strangeness production, propagation and
reabsorption during the heavy-ion collision.

For further details of the model and the explicit parametrizations of
the various reaction cross sections we refer the reader to Refs.
\cite{Rudy,24,Ca97,Brat97}, respectively.

\section{Recoil momentum distributions}

Since in the transport calculations the four-momenta of all hadrons are
propagated in time we are capable to extract for  events,  leading to
hypernucleus formation,  the average properties of the residual
hypernucleus.  For this purpose we compute as a function of time those
particles (essentially nucleons) that have left the residual heavy
fragment at position {\bf R}, i.e.
\begin{equation}
\label{e2}
|{\bf r}_i - {\bf R}|  \geq R_A + 2 fm,
\end{equation}
where $R_A = 1.2 fm \  A_t^{1/3}$ denotes the radius of the target with mass 
number
$A_t$. Now let the number of particles emitted be $N_p(t)$. For each parallel
ensemble we can then evaluate the fragment's average mass number, its
excitation energy, three-momentum and angular momentum  by exploring the
conservation of total energy, mass number, momentum and angular momentum:
$$< E ^* >(t) = E_{tot} - \sum_{j=1}^{N_p(t)} \sqrt{p^2_j + M_j^2}
- M_{res} - E_{coul} - M_{\Lambda} + M_N - E_K ,$$
$$ < A_F >(t) = A_t + 1 - N_p(t) , $$
$$< {\bf p } >(t) = {\bf P}_{tot} - \sum_{j=1}^{N_p(t)} {\bf p}_j(t) ,$$
\begin{equation}
\label{e3}
< {\bf L } >(t) = {\bf L}_{tot} - \sum_{j=1}^{N_p(t)} {\bf r}_j(t)  \times
{\bf p}_j(t) .
\end{equation}
In eq. (\ref{e3}) $M_{res}$ denotes the mass of the 'residual nucleus',
$E_{coul}$ stands for the Coulomb energy between the emitted particles
and the 'residual nucleus', while $E_K$ represents the total energy of
the $K^+$-meson.

All quantities in (\ref{e3}) depend explicitly on time t due to the
continuous evaporation of particles from the final compound system.
Since we will follow the further decay chains by statistical model
codes, the actual transition time for the connection of the BUU and the
statistical model calculation is of no significance as long as the
system has left the nonequilibrium phase of the reaction and achieved
statistical equilibrium.  We have checked that it is sufficient  to
trace the history of each ensemble of events within BUU up to 150 fm/c
\cite{Rudy2}.

As an example for the momentum distribution of hypernuclei we show in
Fig. 2 the computed distributions in beam direction ($dN/p_z$) and
perpendicular to the beam ($dN/dp_x$) for p + $^{238}U$ at $T_{lab}$ =
1.5 GeV from the BUU calculation (dashed lines) as well as after
particle evaporation (solid lines) computed via PACE2
\cite{PACE,Rudy2}.  We note that the particle evaporation only broadens
the momentum distribution somewhat, but does not change the average
momentum per particle in the beam direction.

\vspace*{5mm}
\psfig{figure=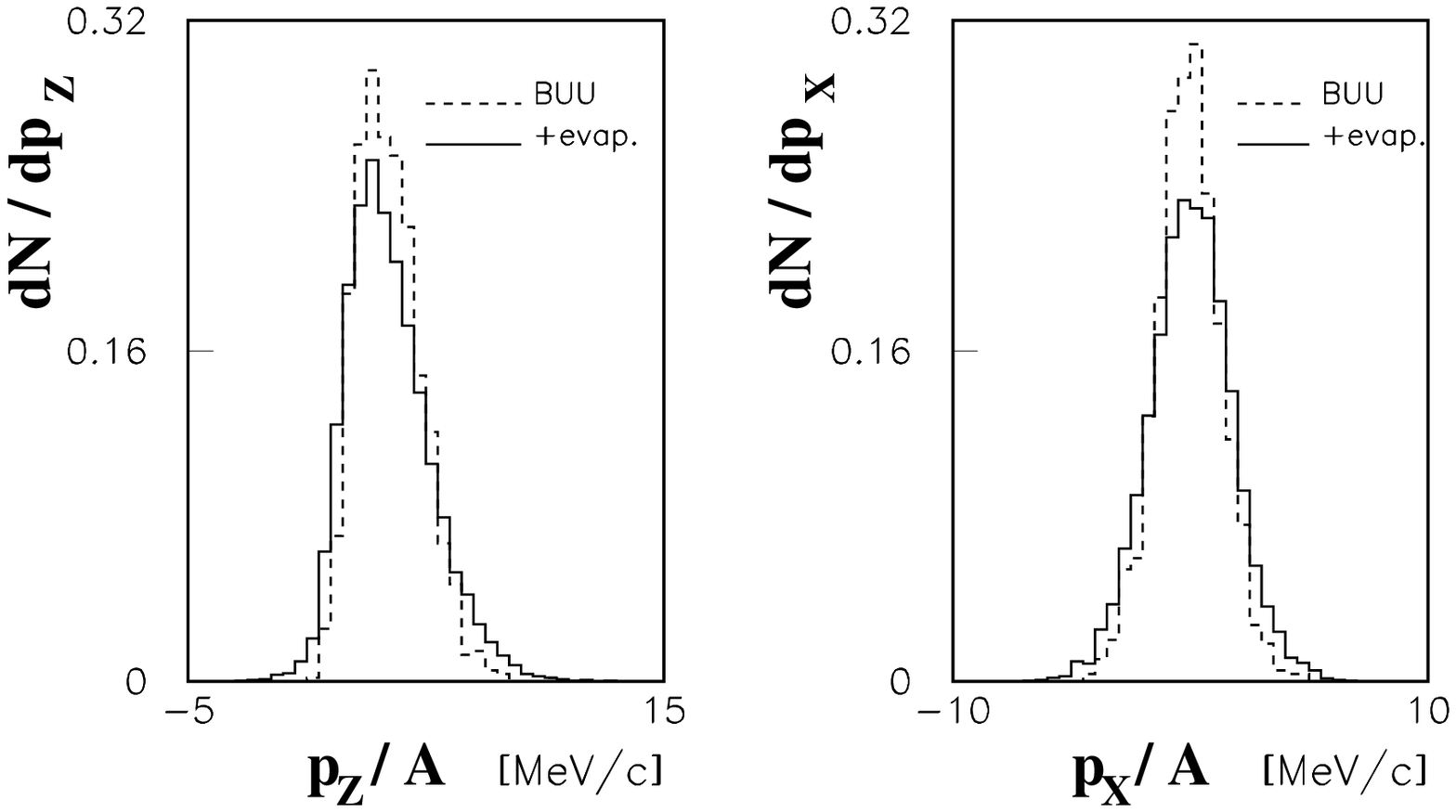,width=15cm,height=8cm}
\vspace*{3mm}
\noindent
{\bf Fig. 2:} Longitudinal momentum ($p_z/A$) and transverse momentum
($p_x/A$) distribution of $\Lambda$-hypernuclei at $T_{lab}$ = 1.5 GeV for
p + $^{238}U$ with evaporation (solid histograms) and without evaporation
(dashed histograms).

\vspace{1cm}

Since the computed momentum distributions enter as an important
ingredient in the experimental analysis \cite{Ohm,Jar} based on the
recoil shadow method, we have to investigate the accuracy of the
transport calculations with respect to the momentum transfer in
proton-nucleus reactions.  In this respect we show in Fig. 3 the
longitudinal momentum distribution of the residual nuclei from the BUU
calculation (solid histograms) - without gating on hypernuclei - for p
+ $^{238}$U at $T_{lab}$ = 475 MeV, 1.0 GeV, 1.5 GeV, and 2.9 GeV in
comparison to the data of Fraenkel et al. \cite{Fraenkel} (full
squares) and Kotov et al. \cite{Kotov}. The good agreement with the
data in this wide kinematical regime demonstrates the relative accuracy
of the transport approach which should be of the same quality when
gating on events with hypernucleus formation (cf. Fig. 2).

\newpage
\psfig{figure=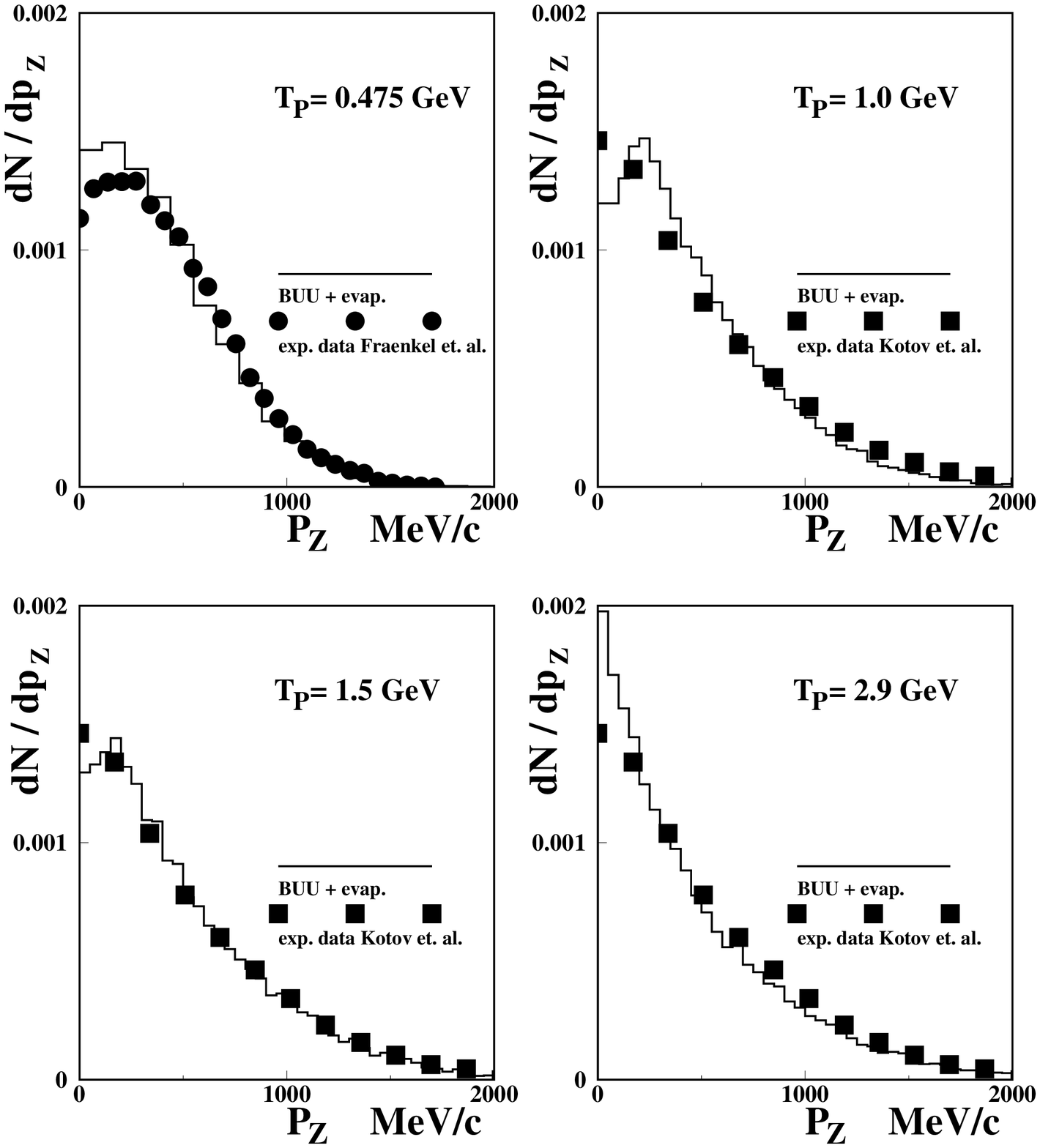,width=15cm,height=18.5cm}
\vspace*{10mm}
\noindent
{\bf Fig. 3:} Comparison of the longitudinal momentum distribution for
p + $^{238}$U at $T_{lab}$ = 475 MeV (lhs. top), 1.0 GeV, 1.5 GeV, and
2.9 GeV from the BUU calculation (solid histograms) with the data from
Fraenkel et al.  \cite{Fraenkel} and Kotov et al. \cite{Kotov} (full
squares).

\section{$K^+$ and $K^-$ production in p + A collisions}
The inclusive production of kaons and hypernuclei in p + A collisions
has been studied at subthreshold energies to a large extend in Ref.
\cite{13,22,Rudy} and does not have to be reviewed here. The novel
aspects are those related to antikaons and their selfenergy or
potential inside heavy nuclei. Since antikaon production has a
threshold of about 2.5 GeV in free nucleon-nucleon collisions, we have
to address the question of kaon production at $T_{lab} \approx$ 2 - 3
GeV, too.

In this respect we first compare our calculations  (solid histograms)
with the experimental $K^+$ spectra for p + Pb at 2.1 GeV from
Schnetzer et al.  \cite{Schnetzer} in Fig. 4 without including any
medium modification of the kaons (in line with the hypothesis of Kaplan
and Nelson \cite{Kaplan} (cf. Fig. 1)). Indeed, the experimental
spectra are described quite accurately as in case of nucleus-nucleus
collisions at SIS energies \cite{Brat97} without any selfenergies such
that a measurement of $K^-/K^+$ ratios at the same bombarding energy
becomes interesting due to its sensitivity to the antikaon potentials.

\vspace*{3mm}\hspace*{15mm}
\psfig{figure=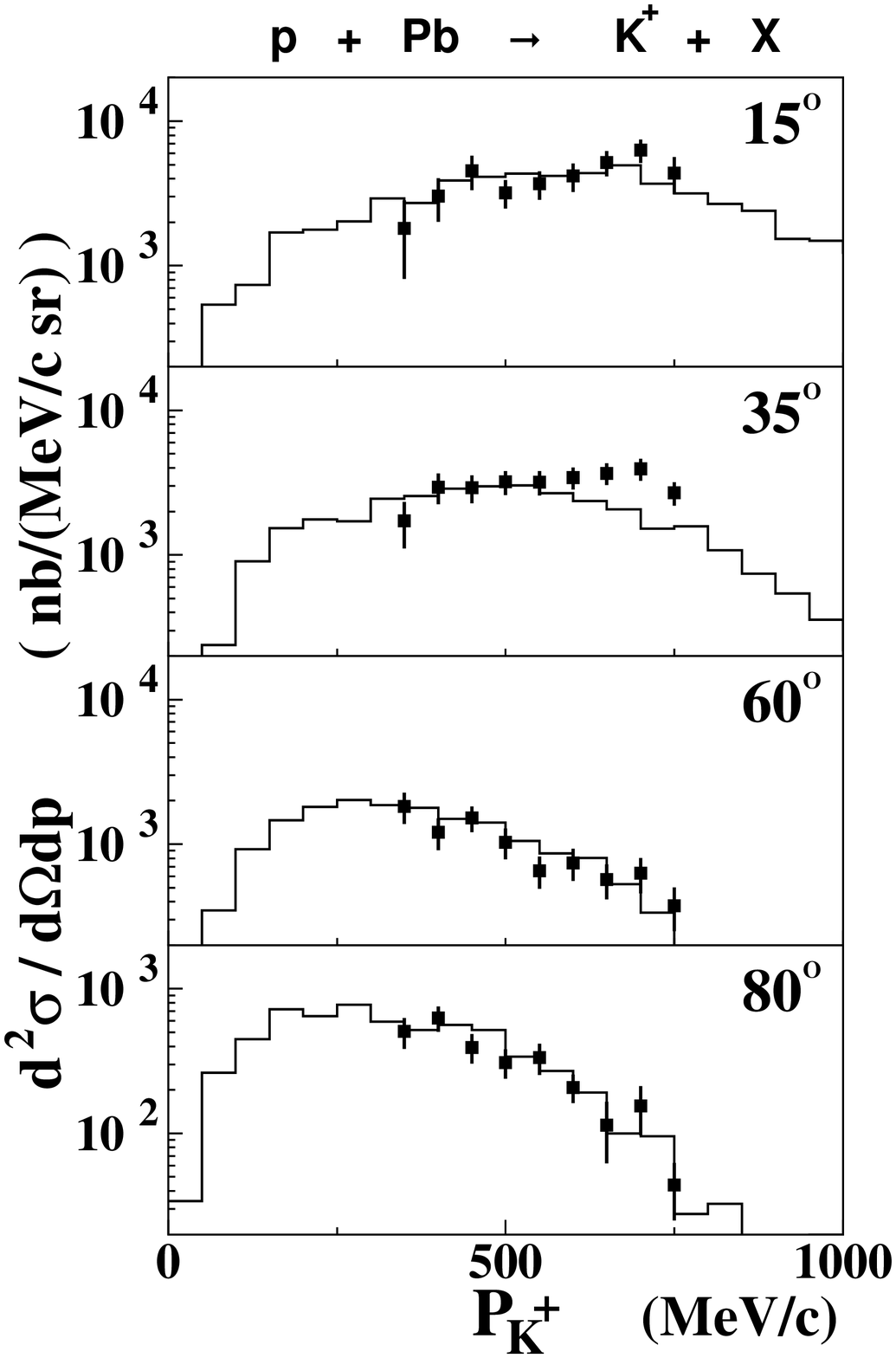,width=9cm,height=13cm}

\vspace*{5mm}
\noindent
{\bf Fig. 4:} The calculated $K^+$ spectra for p + Pb collisions (solid
lines) at 2.1 GeV in comparison to the data from Schnetzer et al.
\cite{Schnetzer} at various angles $\theta_{lab}$ in the laboratory.

\vspace{1cm}

Employing the various production channels for antikaons ($pN
\rightarrow NN K\bar{K}, \pi N \rightarrow N K\bar{K}, \pi Y
\rightarrow N \bar{K}, Y N \rightarrow NN \bar{K}$) as well as antikaon
absorption ($\bar{K} N \rightarrow \pi Y$) from Ref. \cite{Ca97} we
have performed detailed calculations on the $K^+$ and $K^-$ spectra in
p + $^{208}$Pb collisions from 2 - 3 GeV bombarding energy. The first
predictions for the total $K^-/K^+$ ratio are presented in Fig. 5
(integrated over all momenta) as a function of $T_{lab}$ for a bare
antikaon mass (open squares) and an in-medium antikaon mass (full dots)
according to Eq. (5) with $\alpha$ = -0.2 which corresponds to an
attractive antikaon potential of about -100 MeV at $\rho_0$. We find
the $K^-/K^+$ ratio to increase from 2$\times$10$^{-4}$ to
7$\times$10$^{-3}$ in this energy regime for the bare antikaon case;
however, when including the attractive $K^-$ potential the ratio is
enhanced by about a factor of 10 at $T_{lab}$ = 2 GeV and by a factor
of $\approx$ 2 at $T_{lab}$ = 3 GeV.

\vspace*{12mm}\hspace*{15mm}
\psfig{figure=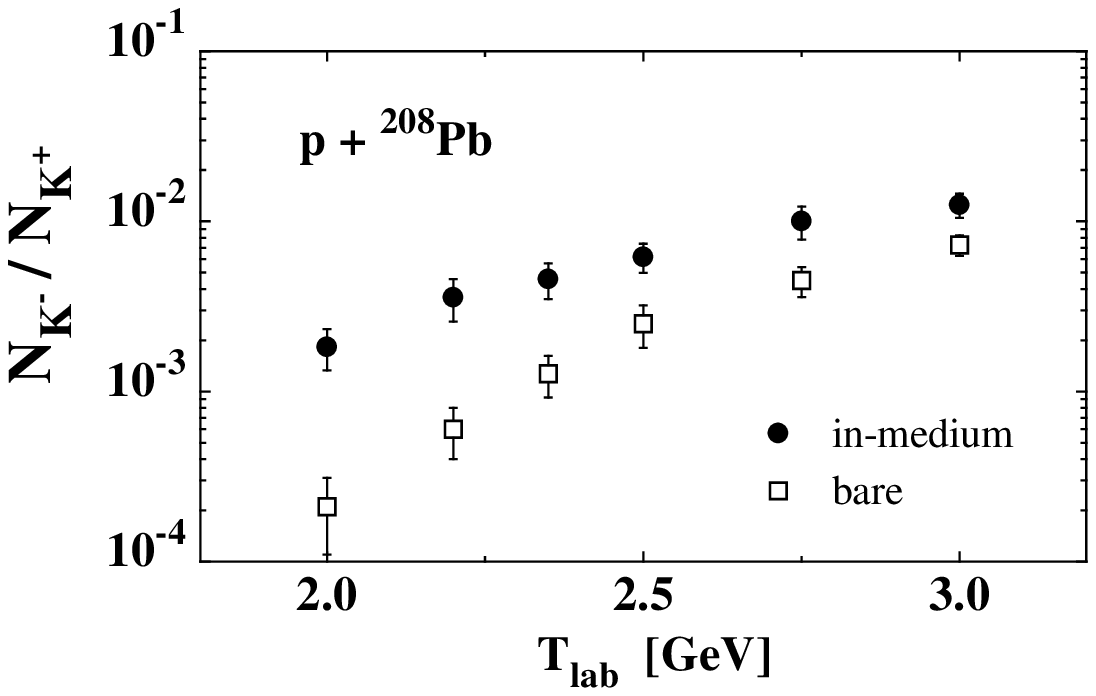,width=10cm,height=7.cm}
\\
\noindent
{\bf Fig. 5:} The calculated $K^-/K^+$ ratio for p + $^{208}$Pb
reactions from 2 to 3 GeV; with an attractive antikaon potential (full
dots); without antikaon potential (open squares).

\section{Summary and conclusion}

In this study we have presented a BUU transport analysis of $K^+ Y$ and
$K^+ K^-$ production in proton + nucleus collisions at COSY energies
employing the elementary production processes from our earlier work
\cite{22,Ca97}. Due to a large fraction of $\Lambda$ hyperons from the
secondary process $\pi N \rightarrow K^+ \Lambda$, which leads to
hyperons with moderate momenta in the laboratory system, the $p +
$nucleus reaction efficiently produces heavy hypernuclei and cross
sections of about a few 100 $\mu b$ are expected for p + $^{208}$Pb at
1.5 - 1.9 GeV \cite{Rudy}, which is in line with present experimental
data so far \cite{Ohm,Jar}.

We have, furthermore, shown that the transport approach also reliably
describes the momentum transfer to the target nucleus in a wide
kinematical regime from 475 MeV to 2.9 GeV for heavy nuclei such that
the computed momentum distributions for hypernuclei, that are needed
for the experimental analysis \cite{Ohm,Jar} based on the recoil shadow
method, are expected to have the same accuracy.

In addition, we have explored the possibility to measure the $K^-$
potential in finite nuclei via their production cross section relative
to $K^+$ mesons. In fact,  p + $^{208}$Pb reactions at 2.0 GeV indicate
an increase of the $K^-$ cross section about a factor of 10 when
including a $K^-$ potential of about -100 MeV at $\rho_0$ (cf. Fig. 1)
in line with the chiral Lagrangians of \cite{Kaplan,waas} and the
analysis in Ref.  \cite{Ca97}. This enhancement should be clearly seen
in the next generation of experiments.

\end{document}